\documentstyle[onecolumn,epsf]{mn}
\newif\ifAMStwofonts
\def\aa#1#2#3{#3, A\&A, #1, #2}
\def\aj#1#2#3{#3, AJ, #1, #2}
\def\apj#1#2#3{#3, ApJ, #1, #2}
\def\jcp#1#2#3{#3, J. Chem. Phys., #1, #2}
\def\mnras#1#2#3{#3, MNRAS, #1, #2}
\def\nature#1#2#3{#3, Nature, #1, #2}

\ifoldfss
  \ifCUPmtlplainloaded \else
    \NewTextAlphabet{textbfit} {cmbxti10} {}
    \NewTextAlphabet{textbfss} {cmssbx10} {}
    \NewMathAlphabet{mathbfit} {cmbxti10} {} % for math mode
    \NewMathAlphabet{mathbfss} {cmssbx10} {} %  "   "    "
  \fi
  \ifAMStwofonts
    \ifCUPmtlplainloaded \else
      \NewSymbolFont{upmath} {eurm10}
      \NewSymbolFont{AMSa} {msam10}
      \NewMathSymbol{\upi}     {0}{upmath}{19}
      \NewMathSymbol{\umu}     {0}{upmath}{16}
      \NewMathSymbol{\upartial}{0}{upmath}{40}
      \NewMathSymbol{\leqslant}{3}{AMSa}{36}
      \NewMathSymbol{\geqslant}{3}{AMSa}{3E}

       \let\le=\leqslant
       \let\ge=\geqslant
    \fi
  \fi
\fi % End of OFSS

\ifnfssone
  \newmathalphabet{\mathit}
  \addtoversion{normal}{\mathit}{cmr}{m}{it}
  \addtoversion{bold}{\mathit}{cmr}{bx}{it}
  \newmathalphabet{\mathbfit} % math mode version of \textbfit{..}
  \addtoversion{normal}{\mathbfit}{cmr}{bx}{it}
  \addtoversion{bold}{\mathbfit}{cmr}{bx}{it}
  \newmathalphabet{\mathbfss} % math mode version of \textbfss{..}
  \addtoversion{normal}{\mathbfss}{cmss}{bx}{n}
  \addtoversion{bold}{\mathbfss}{cmss}{bx}{n}
  \ifAMStwofonts
    \ifCUPmtlplainloaded \else
      \UseAMStwoboldmath
      \makeatletter
      \new@mathgroup\upmath@group
      \define@mathgroup\mv@normal\upmath@group{eur}{m}{n}
      \define@mathgroup\mv@bold\upmath@group{eur}{b}{n}
      \edef\UPM{\hexnumber\upmath@group}
      \new@mathgroup\amsa@group
      \define@mathgroup\mv@normal\amsa@group{msa}{m}{n}
      \define@mathgroup\mv@bold\amsa@group{msa}{m}{n}
      \edef\AMSa{\hexnumber\amsa@group}
      \makeatother
      \mathchardef\upi="0\UPM19
      \mathchardef\umu="0\UPM16
      \mathchardef\upartial="0\UPM40
      \mathchardef\leqslant="3\AMSa36
      \mathchardef\geqslant="3\AMSa3E

       \let\le=\leqslant
       \let\ge=\geqslant
    \fi
  \fi
\fi % End of NFSS release 1

\ifnfsstwo
  \DeclareMathAlphabet{\mathbfit}{OT1}{cmr}{bx}{it}
  \SetMathAlphabet\mathbfit{bold}{OT1}{cmr}{bx}{it}
  \DeclareMathAlphabet{\mathbfss}{OT1}{cmss}{bx}{n}
  \SetMathAlphabet\mathbfss{bold}{OT1}{cmss}{bx}{n}
  \ifAMStwofonts
    \ifCUPmtlplainloaded \else
      \DeclareSymbolFont{UPM}{U}{eur}{m}{n}
      \SetSymbolFont{UPM}{bold}{U}{eur}{b}{n}
      \DeclareSymbolFont{AMSa}{U}{msa}{m}{n}
      \DeclareMathSymbol{\upi}{0}{UPM}{"19}
      \DeclareMathSymbol{\umu}{0}{UPM}{"16}
      \DeclareMathSymbol{\upartial}{0}{UPM}{"40}
      \DeclareMathSymbol{\leqslant}{3}{AMSa}{"36}
      \DeclareMathSymbol{\geqslant}{3}{AMSa}{"3E}

       \let\le=\leqslant
       \let\ge=\geqslant
    \fi
  \fi
\fi % End of NFSS release 2

\ifCUPmtlplainloaded \else
  \ifAMStwofonts \else % If no AMS fonts
    \def\upi{\pi}
    \def\umu{\mu}
    \def\upartial{\partial}
  \fi
\fi

\title{Distortion of Globular Clusters by Galactic Bulges}

\author[H.K. Nordquist et al.]
{
Holly K. Nordquist, Robert J. Klinger, Pablo Laguna,
and Jane C. Charlton\\
Department of Astronomy \& Astrophysics and CGPG
Penn State University, University Park, PA 16802, USA
}
\date{Accepted 1998 December XXX. Received 1998 December XXX}
\pagerange{\pageref{firstpage}--\pageref{lastpage}}
\pubyear{1994}

\begin{document}

\maketitle

\label{firstpage}

\begin{abstract}
One of the external fields that influences the population of globular clusters
is that due to galactic bulges. In extreme situations, perigalactic distances
$r_p \le 100$ pc, globular clusters could suffer total
disruption in a single passage.
A more common scenario is that for cluster orbits with $r_p \ge 200$ pc.
We investigate the effects of tidal forces from a bulge on the shape
of globular clusters for this type of encounters.  We find distortions
characterized by ``twisting isophotes'' and consider the potential for
observability of this effect.  In the Milky Way, a typical globular cluster
must pass within several hundred pc of the center to experience substantial
distortion, and it is possible that this has happened recently to one 
or two present day clusters.  We estimate that this distortion could be 
observed even for globulars in dense fields toward the bulge.  In more 
extreme environments such as giant ellipticals or merger products with newly 
formed globulars, this effect could be
more common, extending out to orbits that pass within 1~kpc of
the bulge center.  This would lead to a substantial shift in the 
eccentricity distribution of globulars in those galaxies.
\end{abstract}


\section{Introduction}

The mechanism by which globular clusters (GCs) are destroyed
has received considerable attention
\cite{tremaine75,chernoff87,aguilar88,chernoff90,long92,capuzzo93,capuzzo97}.
These studies include intrinsic processes like cluster evaporation,
as well as environmental influences such as dynamical
friction and tidal shocking from disks, bulges and supermassive
black holes.
Recently, Gnedin and Ostriker (1997)
considered the effect of these processes on
the present and, by extrapolation, on the initial population
of GCs in the Milky Way.
The conclusion they drew is that between 52\% and
86\% of the current population of GCs in our galaxy
will be eventually destroyed by the combination of evaporation,
dynamical friction and tidal shocking
over the next Hubble time.  They added that
the initial population of GCs was
likely to be substantially larger than the present
one, with the clusters that pass closest to the
galactic center subject to destruction by
bulge shocking.  Moreover, the remnant stars
join the bulge and the halo of the galaxy,
perhaps making a dominant contribution to their
stellar populations.  Murali and Weinberg (1997a) reached
similar conclusions (i.e. a factor of two decrease
in the initial population) considering the global
evolution of the populations of Milky Way disk
and halo clusters.

Observationally, the distributions of GC
shapes have been studied in the Milky
Way \cite{white87} and in nearby galaxies such as
M31 \cite{lupton89}, the LMC \cite{kontizas89},
and the SMC \cite{kontizas90}.  A comparison
of GCs in the Milky Way and M31 \cite{han94} shows that both
galaxies have similar, nearly spherical distributions of
clusters, with axial ratios peaked respectively at $0.95$
and $0.93$.
On the other hand, GCs in the LMC and SMC differ significantly.
Globulars in these galaxies require triaxial
parameters to describe their shapes.
The distribution of axial ratios of GCs in the LMC peaks at 
about $0.85$ and that of the SMC, at $0.73$. Han and Ryden (1994)
explain this situation by a gradual decay of velocity
anisotropy as a consequence of two body relaxation, so that
the older clusters in the Milky Way and M31 are now supported
mainly by rotation.  However, Goodwin (1997) argues that
there is no age/ellipticity relationship for GCs, and that
in fact the stronger tidal field in larger galaxies is the
dominant effect in producing the observed differences.
Severe observed distortions in the
globulars of giant galaxies are rare, but
axial ratio values as low as $0.75$ have been measured
for GCs in the Galaxy.  
Also, the discovery of populations of young GCs in various 
interacting galaxies opens a new laboratory for the study of 
destruction mechanisms 
\cite{holtzman92,whitmore93,oconnell94,whitmore95,meurer95,holtzman96,conti96,watson96,schweizer96}.
For external galaxies and for the Milky Way
observations of distortion typically involve measurement
of ellipticity at the half mass radius of
the GC, but it is also known that the shape changes
with distance from cluster center.

Theoretical inferences about the destruction of GCs
have mostly been considered from the statistical point
of view. In a previous paper \cite{charlton95}, we instead
considered how the destruction event itself proceeds. There we 
reported the outcome of a series of
$N$--body numerical simulations of a single GC
under the influence of the gravitational potential
of a galactic bulge and/or supermassive black hole.
Disk shocking also plays an important role, however,
in the inner parts of galaxies it is the central
spheroid that dominates \cite{aguilar88}.
In our previous paper, we concentrated on computing
the mass lost by the GC as a
function of the distance of closest approach, the bulge
concentration, and the mass of the black hole.
A GC venturing within a distance of a few hundred pc
of the galactic center was completely destroyed in a single passage.
However, it is expected that these ultra-close encounters are relatively
rare.  More typically, the orbit of a GC would pass through a range 
of perigalactic distances for which only a small fraction of
the stars become unbound.  In these types of encounters, the GC
would be noticeably distorted by tidal forces, but would
not be totally destroyed until at least several passages
had taken place.  Many of the clusters on orbits that venture
close to the bulge were destroyed billions
of years ago, however as the GC system evolves there
will be some clusters that reach closer and closer passages.
Several Galactic clusters are presently found within 1--2~{\rm kpc}
of the Galactic center \cite{gnedin97}.

The main goal of this paper is to characterize the
distortions induced in a GC by its interaction
with a galactic bulge. For some cases the effect of
a supermassive black hole is also considered.
Of particular interest is the feasibility of 
finding direct observational evidence of the destruction 
of GCs, both in the Milky Way and in external galaxies.
Grillmair et al. (1995) have determined by star count
analysis that most of the twelve Galactic globular clusters 
in their sample have some degree of distortion of their
surface density contours.  They suggest that this
is the effect of tidal stripping of loosely bound stars
by the gravitational potential of the galaxy.  

As with our previous work,
we base the present study on $N$--body simulations that are
designed both to quantify the nature of the distortions of 
GCs produced specifically by bulge shocking and to
explore the region of parameter space in
which such distortions occur.  With our simulations,
it is possible to address past
and present distortions in the Milky Way and to
consider implications in external galaxies of 
various types.  Specifically, we ask: 1) Under which
circumstances do clusters change shape due
to bulge shocking?, 2) What is the
detailed physical mechanism that gives
rise to shape distortions and how can we
characterize them?, 3) Can these distortions
feasibly be observed in the Milky Way, and
are the present constraints consistent with
the properties of the Milky Way bulge and
GC population?,
4) Where and when are bulge
shocking distortions most easily observed?, and
5) How long do distortions persist and what is the
fate of these distorted globulars?

\section{Tracking the Shape of a Globular Cluster}

The main challenge in a numerical study aimed at
characterizing shape distortions of GCs
is to achieve a satisfactory level of resolution.
Typical GC deformations often only involve
the outer layers of the cluster. Thus, in order to obtain reasonable
observational predictions on shapes of GCs from $N$--body simulations,
it is important to model GCs (systems of $\ge 10^5$ stars)
with at least $10^4$ particles.
This lower bound on the number of particles is necessary
because GCs are highly concentrated objects,
requiring a large investment of computational resources to
simultaneously resolve their cores as well as their
outer layers.
These days, $N$--body simulations with $10^{4-5}$ particles are not
considered extremely demanding; however, it is important to utilize a code
fast enough that parameter space searches are feasible.
We use an $N$--body, parallel, oct--tree
code for this purpose \cite{salmon94}.
With this code, it is possible to carry out selected runs with
each star in the GC represented by its
own particle, and, at the same time, the code's performance allows the
exploration of substantial region of parameter space.
The code does not include the effects of two--body relaxation, but
it should not be important for the single passages through the bulge.

Much of the previous work on the evolution of GCs has been based
upon solution of the Fokker--Planck equation (see \cite{oh92},
\cite{murali97a}, \cite{murali97b}, \cite{gnedin97}, and references therein).
Clearly, this technique is convenient for considering the statistical
evolution of the initial Milky Way GC population, including
both internal (stellar evolution) and external (tidal) effects.
However, recent work by Zwart et al. (1998) questions the general 
applicability of this technique by comparing to an N--body approach,
considering the effects of close encounters.  For some
cluster parameters and initial conditions the Fokker--Planck technique
yields GC lifetimes a factor of ten smaller.
This issue remains open for debate, however, since our goal in
this paper is to focus on observability of GC distortions due
to bulge shocking during a single passage, an N--body approach 
is more appropriate regardless.

A GC experiences forces due to the other stars in the cluster and
due to all the other material in its host galaxy.
We model the {\it background} force of the galaxy by an analytic potential.
A crucial question for this study was the choice of bulge potentials.
There is convincing evidence in favor of triaxial bulges
in the Milky Way and in other galaxies \cite{zhao94,bertola91}.
Near--infrared observations from COBE DIRBE support
a Galactic bulge resembling a prolate spheroid with
1:0.33:0.23 axis ratios \cite{dwek95}.
Orbits of GCs in triaxial potentials can certainly
be quite different from those with spherical symmetry (e.g. box orbits).
For that reason, initially we considered triaxial potentials such 
as the Schwarzschild triaxial potential \cite{zee83}.
However, we found that for a single passage with perigalactic 
distance $r_p \ge 200$pc,
there was no significant difference in the nature and shape of the
deformations between triaxial and spherical potentials
as long as both potentials were comparable in depth and in extent (see below).
As a consequence, information about the triaxiality of the bulge
can only be extracted from study of GC shape distortions in a
statistical sense, and this is beyond the scope of this paper.

Based on the above observations and in order to directly
compare with our previous work, we have chosen to model the bulge
using the Hernquist potential \cite{hernquist90}
\begin{equation}
\phi = \frac{\phi_o}{1+r/a} \, ,
\label{eq:1}
\end{equation}
where $\phi_o = -G M_{bg}/a$, $M_{bg}$ is
the mass of the bulge, and
$a$ is a scale length related to the half-mass radius
$r_{1/2} = (1+\sqrt{2})\,a$.  We considered values of $a=400$~pc and
$a=800$~pc.  This represents a range typical of the
bulges of spiral galaxies. In addition, the mass of the bulge
was taken to be $M_{bg} = 10^5 M_{gc}$ or
$M_{bg} = 10^6 M_{gc}$, where $M_{gc}$ is the mass of the
GC (results are presented in units of $M_{gc}$).
These bulge masses are consistent with a typical GC mass
of $10^{5}$M$_{\odot}$ if the bulge has a mass
of $10^{10}$M$_{\odot}$ or $10^{11}$M$_{\odot}$,
respectively.

The tidal force on a star that is on a line joining the centers of 
the bulge and the GC is
\begin{equation}
F_T \approx \frac{GM_{bg}r_*}{(r+a)^3} \, ,
\label{eq:1.3}
\end{equation}
where $r$ is the distance between the bulge and the GC,
and $r_*$ is the distance from the star to the center of the GC.
Notice that at a fixed separation $r$, the strength of this force 
decreases as $a$ increases.
One can find an estimate of the tidal radius, $r_t$, by equating
expression (\ref{eq:1.3}) to the restoring force from the
GC itself. This yields
\begin{equation}
r_t \approx 46.4\, r_*\left(\frac{M_{bg}}{10^5 M_{gc}}\right)^{1/3}-a \,.
\label{eq:1.4}
\end{equation}
A GC will undergo tidal deformations
if $r_p \sim r_t$. Since observations suggest
that $a \ga  400$pc and since we are not interested in
ultra-close encounters (i.e., we consider only $r_p \ge 200$pc), 
one has from (\ref{eq:1.4})
that the region in the GC subject to distortions is $r_* \ga 13$ pc.
Typical core radii of GCs lie between 0.3 and 10 pc \cite{spitzer87},
thus it is only the {\it halo} of the GC that gets distorted.

To understand the role that triaxiality plays in producing
tidal deformations, let us extend the Hernquist potential (\ref{eq:1})
to a triaxial form by defining
\begin{equation}
\frac{r}{a} = \left ( \frac{x^2}{x_o^2}
+ \frac{y^2}{y_o^2} + \frac{z^2}{z_o^2} \right )^{1/2} \, ,
\label{eq:1.1}
\end{equation}
with $x_o$, $y_o$ and $z_o$ as parameters.
The tidal force along the $x$-axis is
\begin{equation}
F_T(x) \approx \frac{GM_{bg}r_*}{(x+x_o)^3} \, ,
\label{eq:1.2}
\end{equation}
with similar expressions holding for the other axes. From 
Dwek et al. (1995), for the Milky Way bulge, 
$x_o \ga 2$kpc and $y_o \sim z_o \sim 0.5$kpc.
Thus, for most of the halo, tidal deformations along the $x$-axis are
negligible.
This in principle represents an anisotropic tidal field that
could yield a clear signature in the cluster deformations.
However, one has to remember that GCs orbits are not exactly
radial and in general not aligned with the major axis of the bulge.
Since the remaining two axes are quite similar, on average the GC
{\it sees} mostly spherically symmetric tidal forces, thus
it suffices to consider spherically symmetric potentials.

For our simulations, we represent the GC by a King model
with a central potential $W_o = 4$ and a half mass radius
of 10 pc.  This is within the range of observed GCs
in the Milky Way, but on the loosely bound end of this
distribution.  For a more tightly bound GC the disruption
and distortion effects would be smaller.
We consider parabolic orbits with 
perigalactic distances $r_p = 200,\, 400,\, 800,\,$ and 
$1600$~pc.  For bound orbits, the speed at perigalacticon
would be smaller so in this sense our calculations give
a lower limit on the amount of distortion.
The initial location of the globular cluster
was sufficiently far from the center of the galaxy so
that tidal effects at the start of the simulation were
negligible.

\begin{table}
\caption{Parameters of runs considered}
\begin{tabular}{lccccc}
Case & $M_{bg}\, (M_{gc})$ & $a$ (pc) & $r_p$ (pc) \\
\hline
C1 & $10^5$ & 400 &  200 \\
C2 & $10^5$ & 400 &  400 \\
C3 & $10^5$ & 400 &  800 \\
C4 & $10^5$ & 400 &  1600 \\
\\
C5 & $10^5$ & 800 &  200 \\
C6 & $10^5$ & 800 &  400 \\
C7 & $10^5$ & 800 &  800 \\
C8 & $10^5$ & 800 &  1600 \\
\\
C9 & $10^6$ & 400 &  200 \\
C10 & $10^6$ & 400 &  400 \\
C11 & $10^6$ & 400 &  800 \\
C12 & $10^6$ & 400 &  1600 \\
\\
C13 & $10^6$ & 800 &  200 \\
C14 & $10^6$ & 800 &  400 \\
C15 & $10^6$ & 800 &  800 \\
C16 & $10^6$ & 800 &  1600 \\
\end{tabular}
\end{table}

Table 1 summarizes the input parameters for runs.
In this paper we focus on identifying
choices of parameters that induced less than 20\% GC mass loss.
We also conducted runs more appropriate for
GC distortion in giant elliptical galaxies,
considering the effect of superimposing a 
black hole of mass $10^{4} M_{gc}$
on the bulge with $a=800$~pc.  In that case, for the
same $r_p$, considerably more mass could be lost due to the
increase in the depth of the potential.
As a test, Case C3 was repeated with $N=10^5$ to
verify that results are not sensitive to $N$ at this
level.

At each step during the evolution, we monitored the 3D shape of
the GC using the inertia tensor
\begin{equation}
I_{ab} = \frac{1}{N}\sum \frac{x_a x_b}{r^2} \, ,
\label{eq:2}
\end{equation}
where the sum is over all the particles in the cluster and
$a,\, b = x,\, y,\, z$.
Although this method provides useful information about the
development of the deformations, it does not take into account projection
effects.  To connect with observations,
following \cite{pascal95},
we also characterize the GC shape by projecting
the cluster onto an arbitrary plane on the ``sky."
The cluster's ellipticity, $\epsilon$, is then obtained from
\begin{equation}
\epsilon \equiv 1 - \frac{\Lambda_-}{\Lambda_+} \, ,
\label{eq:3}
\end{equation}
where
\begin{equation}
2 \Lambda_\pm = (I_{11}+I_{22})\pm\sqrt{(I_{11}+I_{22})^2
-4(I_{11}I_{22}-I_{12}^2)}
\label{eq:4}
\end{equation}
and $I_{ab}$ ($a,\, b = 1,\, 2$) are the eigenvalues of the 2D
inertia tensor. This ellipticity, of course, depends on the orientation
that is chosen to perform the projection.  Thus we must compute the
probability, $P(\epsilon)$, of viewing the cluster with a given ellipticity.
The procedure for estimating $P(\epsilon)$ is as follows:
Given a snapshot of a GC from an encounter with the bulge,
we define first a coordinate system with its origin at the cluster's
center of mass. At this origin,
we select $51 \times 51$ directions
with solid angle $d\Omega = \sin{\theta}d\theta d\phi$ that cover
uniformly the $(\theta, \phi)$ plane.
The GC is then projected in planes perpendicular to those $51 \times 51$
directions, and its ellipticity is computed. Finally,
$P(\epsilon)$ is obtained from a histogram of these ellipticities.

\begin{figure}
\leavevmode
\epsfxsize=4.2truein\epsfbox{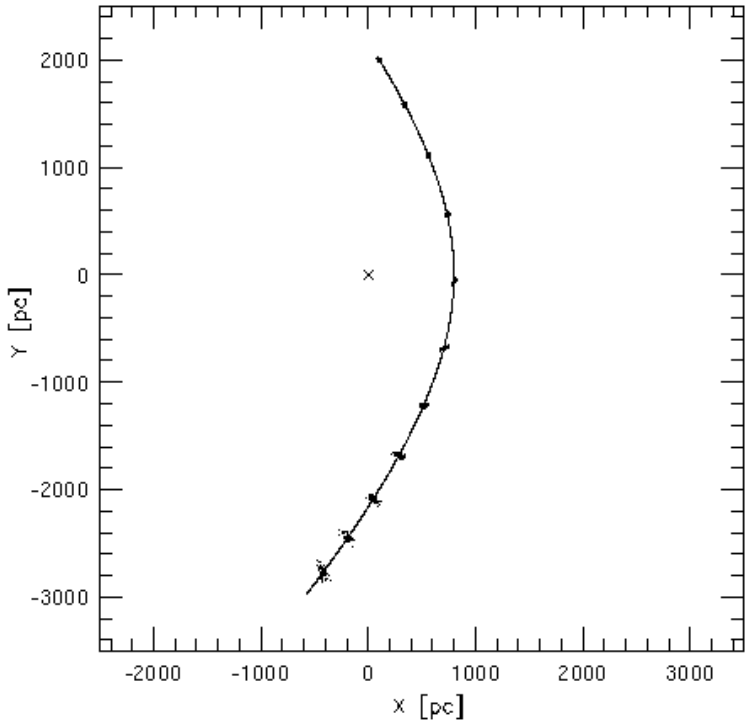}
\caption{Snapshots of a globular cluster passing through
a galaxy bulge, drawn to scale on the orbit.  For this
Case (C3) the bulge has a mass
of $10^5$M$_{gc}$, a concentration of $a = 400$~pc,
and the perigalactic distance of the orbit $r_p = 800$~pc.
The X marks the location of the galactic center.
Each snapshot along the orbit occurs at an interval of
2.3 million years (a unit of time is $4.6 \times 10^4$~yrs).
An expanded view of the clusters for $T=0$, 200, 350, and
500 time units is shown.  These expanded views are not to
scale relative to each other.
\label{fig1}}
\end{figure}

\section{Distorting the Shape of Globular Clusters}

A reasonable starting point to understand the onset of tidal 
deformations of GCs by galactic bulges is to show the outcome of a 
simulation for a typical orbit.
Figure~1 shows the evolving GC along the orbit for case C3,
with blow--up views at selected points. 
The GC has been projected on the orbital plane since distortions
along this plane are dominant for these encounters.
Distortion begins as soon as the GC reaches the tidal radius;
for this case $r_t \sim 528$~pc
for stars at twice the half mass radius of the cluster.
The mechanism behind the distortion of a GC is basically the same as
that for the case of self-gravitating spheres of gas (stars).
That is, the energy required to induce deformation or even
disruption is provided at the expense of the orbital kinetic energy
of the GC.  As mentioned before, and clearly shown in Fig.~1,
the core of the GC remains almost intact.  Only the {\it halo} of the
GC develops a quadrupole distortion that can be spun up via
gravitational torques \cite{rees88}.
Stars closest to the center of the bulge move faster than those
furthest, resulting in the spreading of energy of the stars in
the GC.  This gives rise to the ``twisting''
isophotes apparent in the distorted clusters shown in Fig.~1.

\begin{figure}
\begin{picture}(324,324)(0,0)
\put(81,0){\epsfxsize=5.5in\epsffile{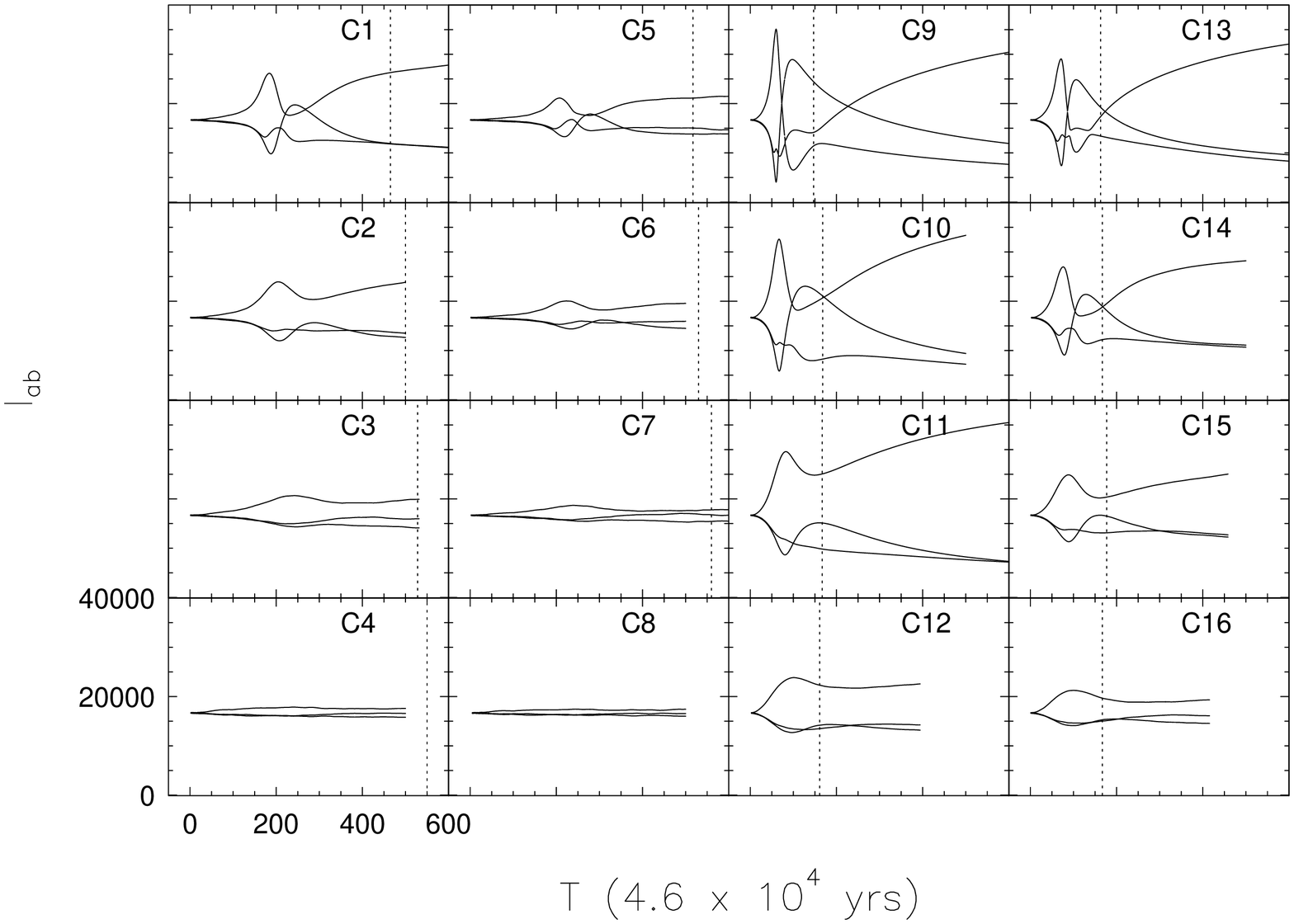}}
\end{picture}
%\vspace{5.5cm}
\caption{Evolution of eigenvalues of the 3D inertia tensor for all cases
in Table~1.  Each column shows a series with increasing perigalactic
approach distance, $r_p = 200$, 400, 800, and 1600~pc for certain
bulge properties (mass $M_{bg}$ and concentration $a$).  From the left
to the right columns we have a) $M_{bg} = 10^5$M$_{gc}$, $a = 400$~pc,
b) $M_{bg} = 10^5$M$_{gc}$, $a = 800$~pc, c) $M_{bg} = 10^6$M$_{gc}$,
$a = 400$~pc, and d) $M_{bg} = 10^6$M$_{gc}$, $a = 800$~pc, respectively.
The vertical dotted lines mark the time, after closest approach, 
at which the GC passes 3~kpc from the galactic center.  One unit of 
time, on the horizontal axis, corresponds to $4.6 \times 10^4$~yrs 
for a $10^5$M$_{\odot}$ GC.
\label{fig2}}
\end{figure}

Figure~2 shows the evolution of the eigenvalues of the 3D inertia
tensor (\ref{eq:2}) for each of the cases in Table~1.
A sense of the degree of deformation is obtained from the relative
size of these eigenvalues. Two general types of deformation can
be detected depending on the value of $r_p$; however, both types lead to
final prolate configurations. For all cases, initially two of the principal
axes (or equivalently eigenvalues) decrease, and the third one
(initially pointing in the direction of the bulge's center) increases.
As the GC continues its orbit (still inside of the tidal radius),
the alignment of the major axis with the line joining the
bulge and GC's centers is no longer possible, in spite of the
induced torques on the GC, which act trying to restore the original alignment. 
Tidal forces halt the growth of the prolate distortion. 
In cases with $r_p = 200$~pc, the situation is 
temporarily reversed, leading to a ``bounce'' with one of the 
short axes becoming the largest (see C9 and C13 in Fig.~2).
The GC eventually settles down into a prolate configuration,
with the major axis in the orbital plane.
A similar, but not as dramatic, prolate configuration is
obtained for larger $r_p$.
In this case the bounce is
absent, with the GC reaching a stable configuration
on a shorter time-scale (about 300 time units or
$\sim 10^7$ yrs after perigalacticon).
The distortion is minimal outside 1~kpc
for $M_{bg} = 10^5 M_{gc}$, but is still substantial
outside this distance if $M_{bg} = 10^6 M_{gc}$.

\begin{figure}
\begin{picture}(324,324)(0,0)
\put(81,0){\epsfxsize=5.5in\epsffile{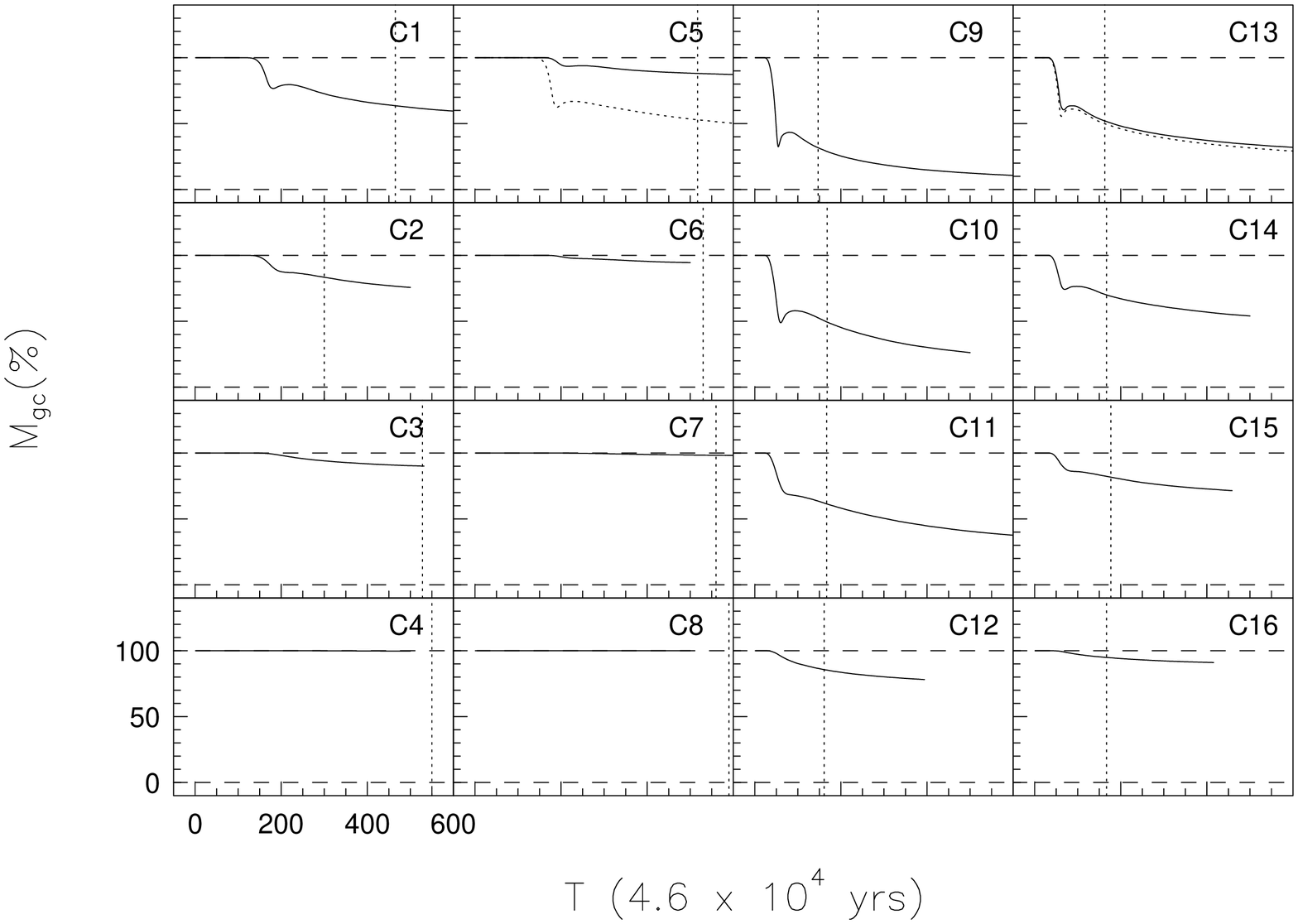}}
\end{picture}
%\vspace{5.5cm}
\caption{Percentage of material that has
become unbound for each of the encounters
in Table~1. Cases are organized as in Fig.~2.  Again, the vertical
dotted lines mark the time at which the GC passes 3~kpc from the center
of the bulge.\label{fig3}}
\end{figure}

The loss of mass during each encounter is displayed
in Fig.~3. The loss of mass steadily increases as $r_p$
decreases, as $a$ decreases, and as $M_{bg}$ increases.
The GC experiences a rapid loss of mass near the perigalactic
distance, followed by a gradual loss over the next
$\sim 10^7$~yrs.  In the cases with $r_p = 200$~pc,
some of the material that was lost in the closest
approach becomes bound to the cluster again during
the ``bounce'' phase, giving rise to an increasing
$M_{gc}$  in Fig.~3 over a short interval of time.
If $M_{bg} = 10^5 M_{gc}$ the GC is destroyed
only if it passes within $\sim 200$~pc of the
galactic center, and distortions could be observed
for $200 \la r_p \la 400$~pc.  For a more massive bulge, approaches
with $r_p = 800$~pc lead to destruction and distorted
clusters will be observed for $1 \la r_p \la 2$~kpc.

\begin{figure}
\epsfxsize=4.2truein\epsfbox{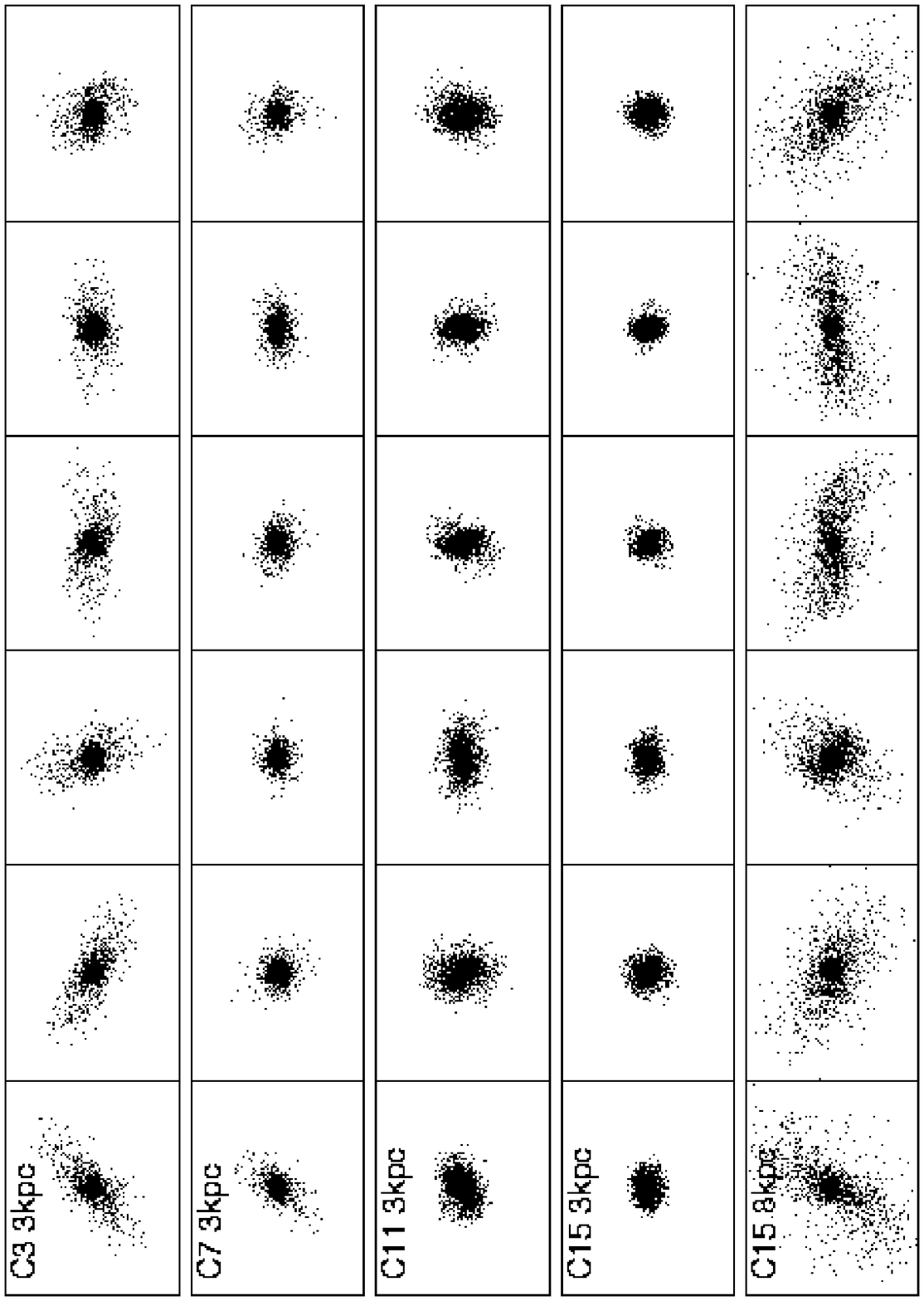}
\caption{Realizations of various randomly oriented clusters.
Each of the first four rows shows six different views of
the cluster from the Case indicated (C3, C7, C11, C15).
This series of cases represents a variety of bulge properties
for a cluster that reaches a perigalacticon distance of
800~pc, viewed after it has reached a point 3~kpc from the
center of the bulge.  C11 and C15 have more eccentric clusters,
however, the distortion in the outer layers is not as apparent
until the cluster has passed further from the galaxy center.
Views of cluster C15 once it has moved to a distance of 8~kpc
are given in the lowest row of panels.
\label{fig4}}
\end{figure}

Even when a cluster is severely distorted, the
observability in any one case depends upon the
viewing angle.  Figure~4 shows four series of
realizations of the randomly oriented clusters
from C3, C7, C11, and C15, all viewed at the time,
after the encounter, when they are 3~kpc from
the galactic center.  These four cases all
have $r_p = 800$~pc, but varying bulge parameters.
Although Case C10 yields the GC with the largest
distortion and thus the larger mean ellipticities,
the ``twisted'' isophotes are more
apparent for Cases C3 and C7.  In fact this is
due to the fact that the GCs in C11 and C15 have
not yet settled into their final states at the
distance of 3~kpc.  The last row in Fig.~4 shows
randomly oriented realizations of the cluster from
C15 at $T = 1.6 \times 10^7$~yrs after reaching the
distance of closest approach, when it is
at a distance of $\sim 8$~kpc.

\begin{figure}
\begin{picture}(324,324)(0,0)
\put(81,0){\epsfxsize=5.5in\epsffile{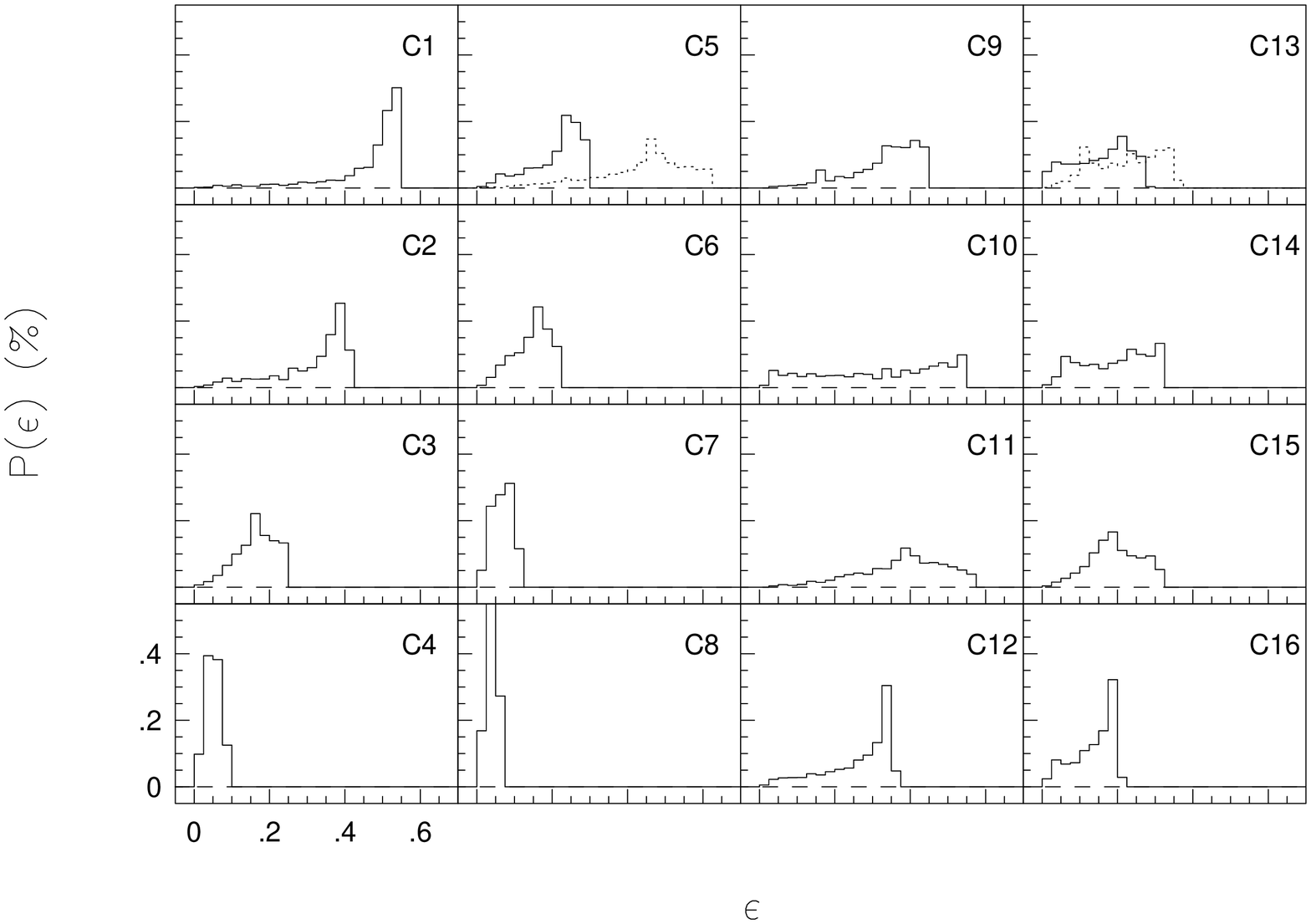}}
\end{picture}
%\vspace{5.5cm}
\caption{Ellipticity probability distribution for each case
in Table~1 when the cluster has reached a distance of 3~kpc
from the center for the bulge.  The organization of cases
is as in Figs. 2 and 3, with increasing $r_p$ from top
to bottom.  The effect of a $10^4$M$_{gc}$ black hole
is illustrated as a dotted histogram for Cases C5 and C13.
\label{fig5}}
\end{figure}

Figure~5 shows the distributions of ellipticities for
each of the cases in Table~1 when the GC is at
a distance 3~kpc from the center of the bulge.
Generally, we see that with increasing $r_p$ 
(along a column in Fig.~5) the eccentricity 
distribution peaks at smaller values.
However, this is not apparent for Cases C9--C11
and Cases C13--15.  In these cases, which experienced
the ``bounce'' described above, the GC has still not
reached a stable configuration when it reaches a
distance of 3~kpc.  Fig.~6 shows eccentricity distributions
for all cases at $T=500$ ($2.3 \times 10^7$~yrs from the
start of the simulation, when the GC was at 2~kpc distance).
These distributions are all narrow and we see a steady
progression with the exception of Case C9.  This is
also the case with the most severe ``bounce''.
The GC eccentricity distribution mean does not reach
the large values we would expect by extrapolating from
the $r_p = 1600$, 800, and 400~pc cases (C12, C11,
and C10).  This is due to the rapid passage of this
GC through the galactic center.  As a result, it does 
not have time to fully respond to the tidal torques
and does not reach as large an ellipticity.

\begin{figure}
\begin{picture}(324,324)(0,0)
\put(81,0){\epsfxsize=5.5in\epsffile{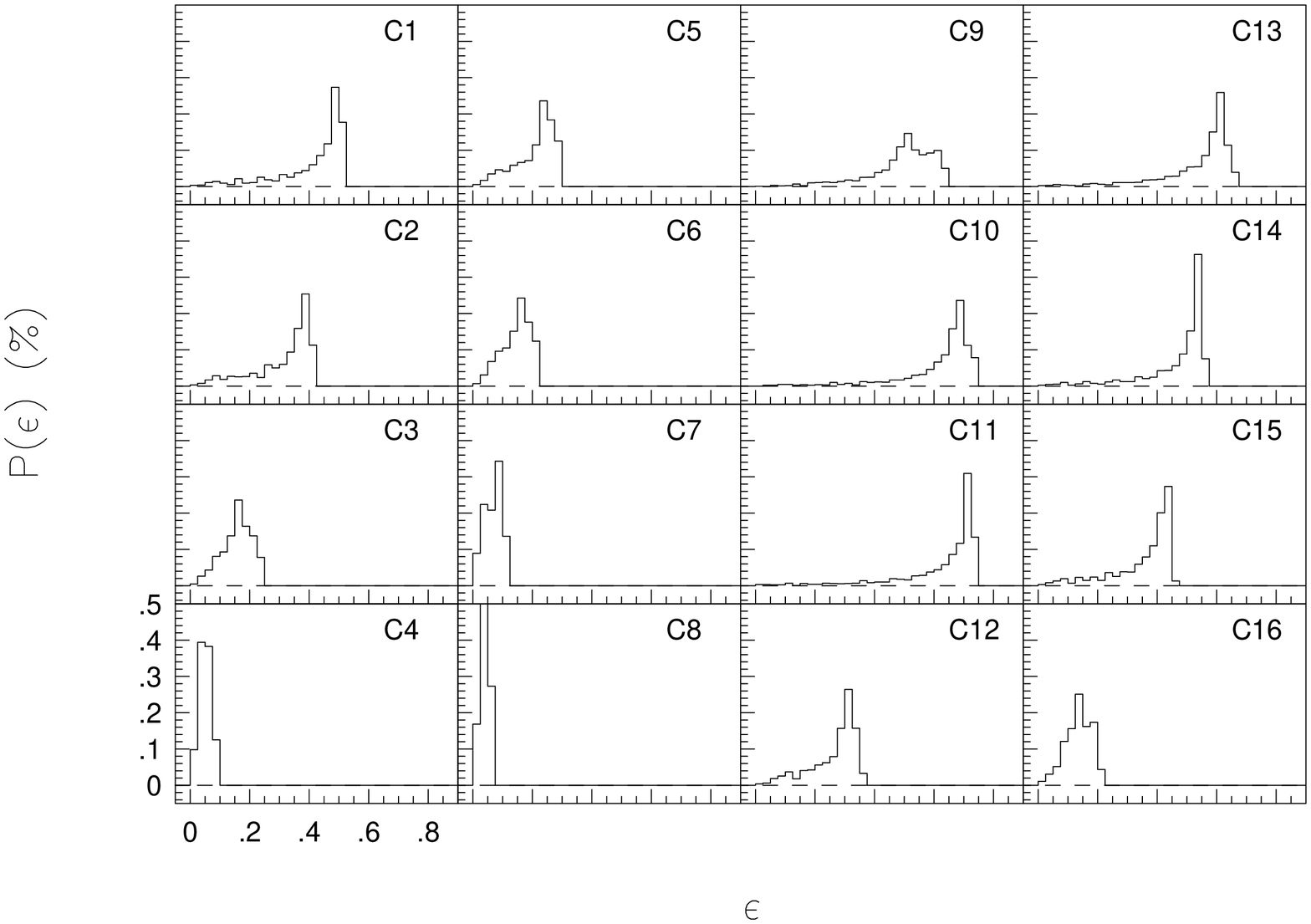}}
\end{picture}
%\vspace{5.5cm}
\caption{Ellipticity probability for each case in Table~1 after
500 time units ($2.3 \times 10^7$~yrs) have elapsed from
the start of each run.  By this time the clusters have reached
a relatively settled state, and a steady progression of
increasing ellipticity with increasing $r_p$ is observed
(with the exception of case C9).  The dotted histograms in
the C5 and C13 panels represent runs with the bulge plus a
$10^4$M$_{gc}$ black hole.
\label{fig6}}
\end{figure}

In giant elliptical galaxies, it seems possible
that the effects of ``bulge shocking'' could be
substantially enhanced by the presence of a
central supermassive black hole.  To address
this specific possibility, we added
a central $10^4$M$_{gc}$ black hole
to runs C5 and C13 ($a = 800$~pc and $r_p = 200$~pc
for both, but with $M_{bg} = 10^5$ and $10^6$M$_{gc}$,
respectively).  The results for mass loss
and for the ellipticity distributions are shown
as dotted lines in Figs.~3, 5, and 6.
In the case with the less massive bulge (C5), the black hole
leads to significantly larger distortions, but the bulge
tidal effect dominates in the case with $M_{bg} = 10^6$M$_{gc}$.

\begin{figure}
\epsfxsize=4.2truein\epsfbox{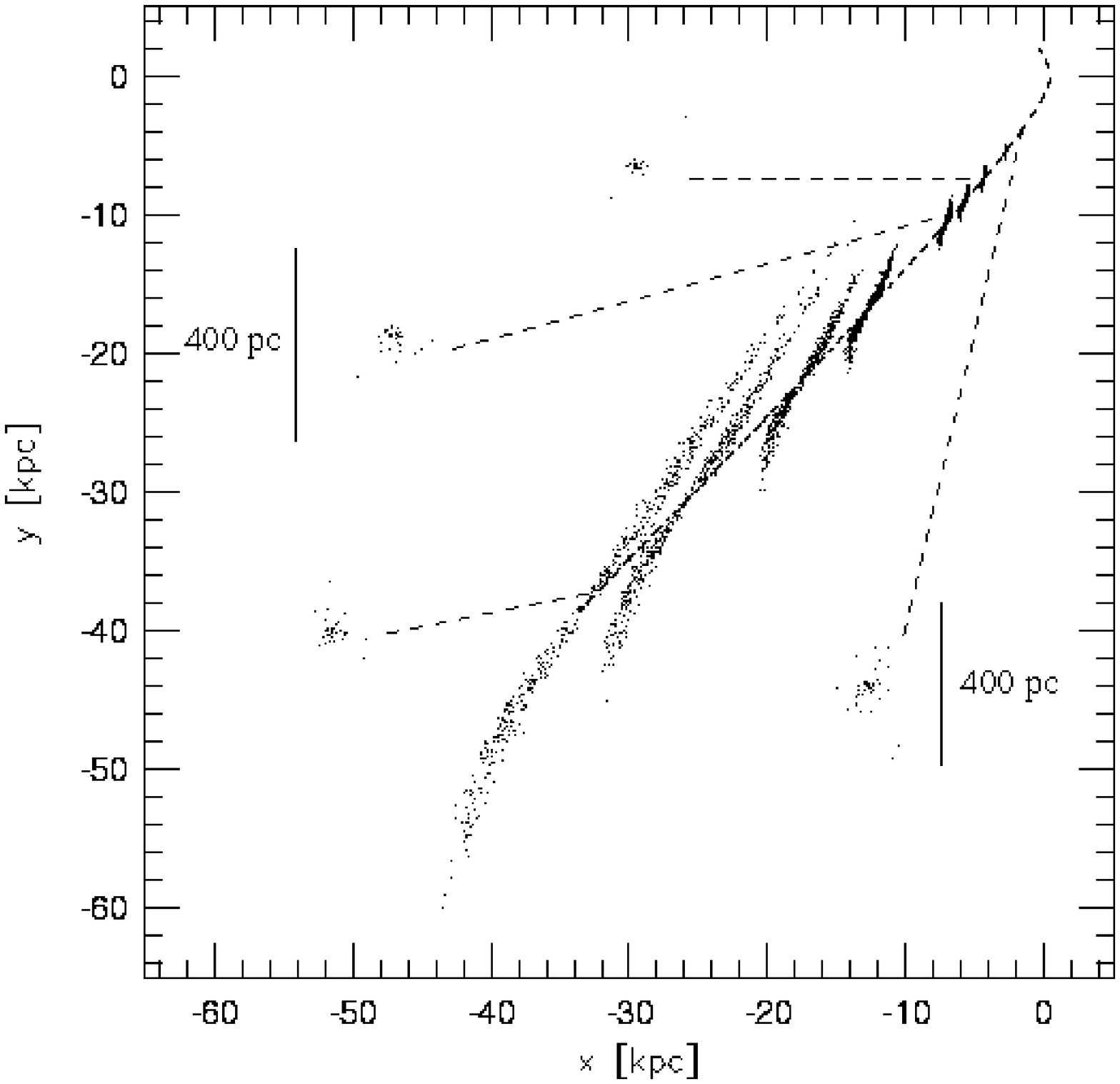}
\caption{Orbit of the Case 2 globular cluster
showing 35000 timesteps ($1.6 \times 10^9$~yrs), and thus
represented by only 10000 particles.  The central regions of
the clusters are shown expanded to squares 400~pc on their
sides at 1000, 3000, 5000, and 35000 timesteps.  Note
that the central region of even the final cluster
appears as a normal globular.  Observable distortions
are no longer apparent at 3000 timesteps.  For
simplicity we show a parabolic orbit, but of course a
cluster will realistically orbit back toward the center
of the Galaxy.
\label{fig7}}
\end{figure}

We have found that, for a range of bulge properties,
there is a fairly large range of perigalactic distances
over which a GC is distorted but not destroyed.
Figure 7 illustrates the evolution of such a cluster
as it continues along a parabolic orbit back out into
the galaxy.    The Case 2 cluster simulation is
shown, but with only 10000 particles, so that its evolution
can be traced for more than 1 billion years.
The distortion is apparent at 1000 time units, corresponding
to $4.6 \times 10^7$~yrs, but is not pronounced at 3000 time
units.  If no further shocking occurs the bound
particles remain behind as a globular cluster while,
within hundreds of millions of years, the outer layers
spread out into a stream extending through the galaxy.
Once dispersed, this stream of stars joins the population
of halo stars.

\section{Are Tidal Deformations Observable?}

We find that GCs can be substantially
distorted, without being destroyed, by the process
of bulge shocking.  Here we address the question of
whether the predicted distortions of the GC halo can
practically be observed in the Milky Way and in other
nearby galaxies.

\subsection{Globular Cluster Distortion in the Present Day Milky Way}
As demonstrated by Gnedin and Ostriker (1997), the rate of
destruction of GCs and the importance of
bulge shocking to their evolution were significantly
larger in the past.  Although the GC
population in the Milky Way is likely to decrease by a
factor of two over the next Hubble time, the probability
of observable bulge shocking at present in the Milky
Way is somewhat difficult to assess.  We use our simulation
results to consider this issue.

The clusters most likely to have observable distortions in
the Milky Way are those that have passed near to
the Galactic center.  There are several clusters observed within
2~kpc, and these are the ones with the highest probability
of having a bulge shocking effect.  As examples, we consider
the clusters NGC6293 (at $R_G = 1.2$~kpc) and
NGC6440 (at $R_G = 1.4$~kpc).  The bulge of the Milky Way
has mass $\sim 10^{10}$M$_{\odot}$ and softening
parameter $a \sim 700$~pc \cite{dwek95}, if the
distribution of light is fit to a Hernquist potential.
With these parameters (closest to the series C5--C8), 
a GC passing within a few hundred pc of
the Galactic center undergoes substantial
distortion of its outer layers.
The outer distorted regions of such a GC
has about 10--100 stars per square arcminute at the
distance of NGC6293 or NGC6440.  In order to determine
whether such distortions are observable, we consider
whether it is possible to distinguish the GC
stars from the background of bulge and disk
field stars in the Galaxy.  The expected backgrounds
in the directions of these two clusters have been
calculated using a Galaxy model constructed by Hunsberger
(private communication).
For GCs $\sim 8$~kpc away from us the range of observed 
apparent magnitudes of their stars is about 12--20.
For NGC6293, at $l=357.6\deg$ and
$b=7.8\deg$, we expect 24 stars per square arcminute
from the bulge and 23 stars per square arcminute from
the disk in this magnitude range.  For NGC6440, at
$l=7.7\deg$ and $b=3.8\deg$,
the bulge contribution is only 9 per square arcminute,
as compared to 21 per square arcminute from disk stars.
In either case the stellar densities in the outer,
distorted regions of the cluster are 10--100
stars per square arcminute, comparable to these
background levels, and should be observable.  Further
information could be extracted by considering
the colors of the stars since a different distribution
is expected for the cluster and for the Galactic
background.

We conclude that the main factor in assessing whether bulge
shocking should be observable in one or more present--day
Milky Way globulars is whether any of the clusters
actually has passed within $\sim$400~pc of the bulge
center.  (The GC that we have chosen is realistic, but fairly 
loosely bound.  The distances at which distortion and destruction
become important would be smaller for a more tightly bound
cluster.)  The key period for observation of the tidal
distortion effect is while the cluster is still within a couple of
kpc of the Galactic Center.  Figure 7 showed that the
distortions do not persist once the cluster has passed
out into the galaxy.  A search for the distortion signature of 
bulge shocking can be used to place limits on the orbits
of present day globulars.  It is expected that most
clusters that passed regularly within the bulge on their orbits 
were destroyed many billions of years ago.  However, three of the
25 globular clusters for which proper motions were measured
were found to have orbits that will bring them within
1~kpc of the Galactic Center, one of them within 300~pc
\cite{dauphole96}.
A careful analysis of the shapes of globulars close to the 
Galactic Center is therefore of interest.

Grillmair et al. (1995) have mapped
the surface density distributions of 12 Galactic clusters
over a range of Galactocentric distances, using star counts
and aided in selection by color--magnitude diagrams.
They present a comparison of one of their most distorted
observed clusters, NGC 7089, to results from their own N--body 
simulations observed in similar manner to the data, and
find agreement.  We have performed a similar analysis and
present in Fig.~8 contours of smoothed pixel maps for six of the 
clusters illustrated in Fig.~7.  These were constructed by
observing each of the oriented clusters from a distance of
8~kpc, and assigning the particles to pixels that were
$1.56 \times 1.56$ square arcminutes, i.e. $3.6 \times 3.6~{\rm pc}^2$.
Contours were drawn by first applying a Gaussian smoothing filter with
$\sigma = 3$~pixels.  Pixels near the center of the image, those
with values larger than the threshhold of 1000 stars per pixel, were
reset to 1000 so that the central structure did not overwhelm the
outer contours in the smoothing process.  The outer contour is
drawn at 2 stars per pixel and the inner contour at 100 stars per 
pixel.  These maps were produced using the {\it imsmooth} and
{\it contour} routines in IRAF\footnote{IRAF is
distributed by the National Optical Astronomy Observatories, which are
operated by AURA, Inc., under contract to the NSF.}.

\begin{figure}
\leavevmode
\epsfxsize=4.2truein\epsfbox{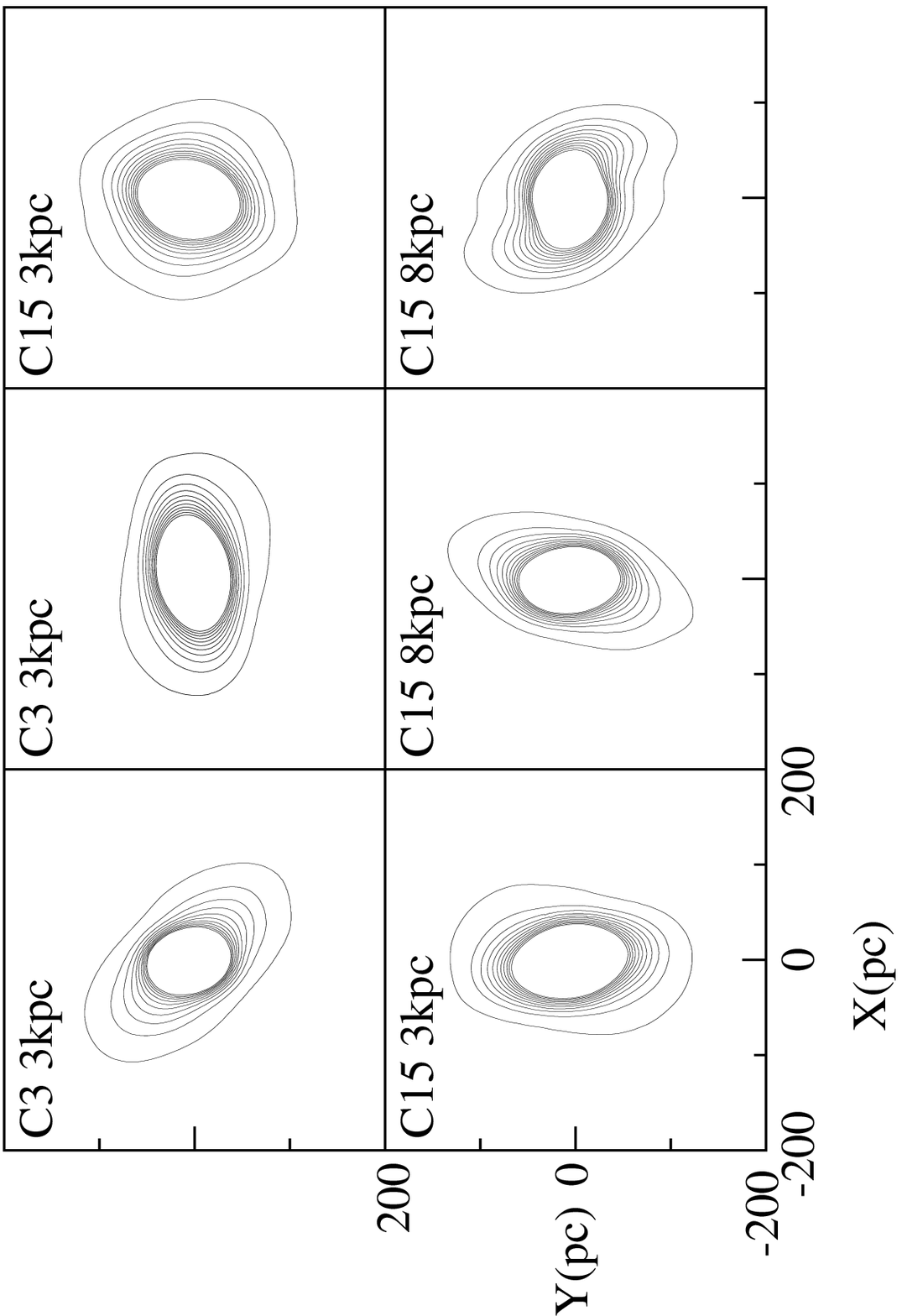}
\caption{Contours for a smoothed pixel map of some representative
clusters.  Each pixel is 1.56{\arcmin} on a side, corresponding
to 3.6~pc for a cluster viewed from a distance of 8~kpc.
The map is smoothed with a Gaussian kernal with $\sigma =3$~pixels
after the central region pixels are set to a maximum of 1000 so
as not to excessively smooth the important outer layers.
The ten contours range from 2 to 100 stars per pixel for
the smoothed map.
The six panels correspond to the following cases
from Fig.~4: realizations 2 and 4 of C3 at 3~kpc, 3 and 4 of
C15 at 3~kpc, and 3 and 4 of C15 at 8~kpc. 
\label{fig8}}
\end{figure}

These parameters used to produce the contours in Fig.~8
are similar to those used in producing the surface density
maps for the 12 Galactic globulars presented by Grillmair 
et al. (1995).  The resulting maps of our similated
GCs are somewhat similar to some of the observed
GCs, suggesting that perhaps tidal distortions have
already been observed in our galaxy.
However, we should note that the Grillmair clusters are all 
at galactocentric distances larger than 8~kpc.  If the tidal 
distortions are in fact due to bulge shocking as we have
discussed, this implies that these clusters have passed
at least within 1~kpc of the Galactic center which is unlikely
for the majority of the twelve.

\subsection{Globular Cluster Distortion in Other Galaxies}

As discussed in the introduction, galaxies of various
types at different stages of evolution show different
distributions of ellipticities \cite{han94}.  In our Galaxy and
M31 there has been time for two--body relaxation to
destroy velocity anisotropies.  We would predict that
distortion should occur in many young galaxies
due to bulge shocking.
Depending on the initial distribution of orbits of GCs
and on their spatial distribution, the early
ellipticity distribution of clusters could be skewed
to low values due to this process.

Some other environments are more likely to currently be 
active sites of extensive bulge shocking.  We repeated runs C5 and
C13 to explore the distortions expected for a parabolic 
orbit passing within 200~pc of the center of an elliptical
galaxy containing a supermassive black hole.
Depending on the orbital distribution, in a young
elliptical with a large GC population this situation may be
more likely than it is for a GC to pass within
400~pc of the Galactic Center.
In general, the initial distribution of clusters could
be spatially concentrated toward the galaxy center,
so that many orbits are subject to bulge shocking.
In a young galaxy with close to its initial distribution
of clusters, we expect many to have the ``twisting
isophotes'' distortions that we describe here.

The merger products of recent interacting giant galaxies
perhaps provide the most ideal laboratory for direct
detection of the destruction of GCs.  As mentioned
in the introduction, in
several of these galaxies more than one thousand
young star clusters are found, with masses somewhat
larger on average than globulars \cite{whitmore95}.
These clusters have properties consistent with
the predecessors of GCs.  We predict
that a number of the globulars in merger products
should exhibit the unique signature of distortion
by bulge shocking.  Globular clusters can just be
resolved at the distances of these interacting pairs
and distortions of their outer layers would be impossible
to observe.  However, the GCs produced by more extreme cases of
bulge shocking (which should be more common in this
environment than in the Milky Way) will have large
eccentricities even at the half mass radius and should
have larger effective radii.  Both relatively gentle
and gradual tidal distortions and more dramatic destructive
shocking must play a role in shaping the distribution of
new globulars.  A statistical study of the
globular cluster distribution as a function of the age 
of the merger product would
provide constraints on its evolution.  Some of the
massive young star clusters in the merger product
NCG 1569 seem to have relatively high ellipticities
\cite{oconnell94}.  Furthermore, some of the young inner
clusters in NGC 7252 have larger effective radii than
any outer clusters in the same galaxy \cite{miller97}.
Constraints based on studies
of the ellipticity distributions of galaxies of various
types could lead to a better understanding
of the formation and evolution of globulars in the
Milky Way and in general.

\section*{Acknowledgments}

We are grateful to M. Warren for making available his Tree-code.
We also thank S. Hunsberger for providing the code used to estimate
the Galactic background of disk and halo stars.
This work was partially supported by NSF 93-57219 (NYI) grant to PL
and by NSF grant 95-29242 awarded to JCC.

\label{lastpage}
\end{document}